\documentclass[prd,superscriptaddress,onecolumn,showpacs,11pt,msmath,preprintnumbers,showkeys,amsmath]{revtex4}
  \usepackage{wrapfig,rotating}
  \usepackage{graphicx,epsfig,color}



  \usepackage{graphicx}
  \usepackage{dcolumn}
  \usepackage{bm}

  \def\beq{\begin{equation}}
  \def\eeq{\end{equation}}
  \def\beeq{\begin{eqnarray}}
  \def\eeeq{\end{eqnarray}}

\def\bas{\bar{\alpha}_s}

  \def\as{\alpha_{\mbox{\rm\scriptsize s}}}
  \def\lrang#1{\left\langle#1\right\rangle}
  \def\cO#1{{\cal{O}}\left(#1\right)}

  \def\2GPD{$_2\mbox{GPD}$}
  \def\GeV{{\rm Ge\!V}}
  \def\MeV{{\rm Me\!V}}
  
  \def\12{$1\otimes 2$}
  \def\22{$2 \otimes 2$}

  \def\cN{{\cal{N}}}
  \def\Qsep{Q_{\mbox{\rm\scriptsize sep}}}
  \def\Qsep2{Q^2_{\mbox{\rm\scriptsize sep}}}
  
\begin{document}

\title{Rapidity distribution of particle multiplicity in DIS at small $x$}

\author{
B.\ Blok$^{1}$, Yu.\ Dokshitzer$^{2}$ and M.\ Strikman$^{3}$
\\[2mm]
\normalsize $^1$ Department of Physics, Technion -- Israel Institute of Technology, Haifa, Israel \\
\normalsize $^2$ CNRS, LPTHE, University Pierre et Marie Curie, UMR 7589,  Paris, France \\
{\small On leave of absence: St.\ Petersburg Nuclear Physics Institute, Gatchina, Russia}
\\ \normalsize $^3$ Physics Department, Penn State University, University Park, PA, USA
}

\begin{abstract}

Analytical study of the rapidity distribution of the final state particles in deep inelastic scattering at small x is presented.
We separate and analyse three sources of particle production: fragmentation of the quark-antiquark pair,
accompanying coherent soft gluon radiation due to octet color exchange in the t-channel, and fragmentation of gluons that form parton distribution functions.

Connection to Catani-Ciafaloni-Fiorani-Marchesini (CCFM) equations and the role of gluon reggezation are also  discussed.
\end{abstract}

\maketitle
\thispagestyle{empty}

\section{Introduction}

Perturbative QCD approach successfully describes internal structure of quark and gluon jets produced in $e^+e^-$ annihilation into hadrons.

The final states of hard lepton--hadron and hadron--hadron collisions have a more complicated structure.
In this case, apart from jets formed by the partons originating from the underlying hard scattering,
fragmentation of the initial state hadron(s) also contributes to the particle yield.

Really,
multiparticle production in hard processes is driven by radiation and successive cascading of relatively soft gluons.
Soft bremsstrahlung is subject to coherence effects. Analysis of such effects in jets has led to ``angular ordering'' (AO) as means of organizing parton multiplication in terms of probabilistic time-like cascades~\cite{AOjet}.
For the space-like case the corresponding problem was addressed and solved by L.~Gribov et al in \cite{GDKT, GDKT1} and independently by M.~Ciafaloni~\cite{Ciafaloni}.

It was shown that in order to formulate the probabilistic picture of the gluon radiation caused by fragmentation of a space-like parton ensemble, one has to impose AO as well, similar to the time-like jet evolution case.
This result was further
developed by S.~Catani, F.~Fiorani and G.~Marchesini, and laid ground for the CCFM scheme for generating final states of small-$x$ DIS processes in accord with the perturbative QCD~\cite{CCFM}.

\smallskip

In this letter we study pseudorapidity (angular) distribution of particles produced in small-$x$ DIS processes.

Similar to \cite{GDKT} where the inclusive energy spectra of final state particles were derived, we find three essential contributions to the answer. The first one ($dn^{(1)}$) originates from fragmentation of the struck quark and its partner antiquark (at small $x$ it is a sea $q\bar{q}$ pair that is hit by an incident lepton).
This contribution dominates the particle yield at large momentum transfer $q^2=-Q^2$.
Two more contributions, formally subleading but rather important, are due to the underlying space-like gluon cascade that produces the quark pair. One of them ($dn^{(2)}$) is driven by coherent soft gluon radiation caused by octet $t$-channel color exchange, and the last one ($dn^{(3)}$) --- by fragmentation of relatively hard (energetic) gluons that determine the hadron structure functions.

Section \ref{sec:1} is devoted to derivation of corresponding analytic expressions for the spectrum of final particles.
In Section  \ref{sec:2} we present numerical results for pseudorapidity distribution in DIS for different values of $x$ and $Q^2$, both academic and realistic.

\section{Approximation and main contributions \label{sec:1}}

 {\em Collinear approximation}\/ allows one to construct a probabilistic QCD cascade picture of multiparticle production and, in particular, to separate initial and final state radiation. Selecting and resuming contributions in which each power of $\as$ is accompanied by a logarithmic integration over parton transverse momentum, gives rise to the Leading Logarithmic  Approximation (LLA).

Because of the double-logarithmic nature of gluon radiation, the true perturbative expansion parameter for observables like  {\em particle multiplicity}\/ turns out to be not the QCD coupling $\as$ itself, but rather $\sqrt{\as}$. So, the next-to-leading logarithmic approximation (NLLA) effects are down by $\sqrt{\as}$ as compared with the LLA, etc.

The general NLLA formulae describing the sub-jet structure of the DIS final state in terms of generating functionals were presented in
the paper that has introduced the $k_\perp$--clustering algorithm for jets in DIS and hadron--hadron collisions~\cite{CDW92}.

This approximation, however, turns out to be insufficient to access the structure of the fragmentation of a target proton in DIS.
This region provides a $\cO{\as}$ fraction of the total particle yield, and therefore is formally of the NNLL nature. At the same time, this kinematical region widens and becomes important for small values of Bjorken $x$.
In \cite{GDKT} it was demonstrated that the perturbative QCD analysis could be carried out, and NNLA expressions could be derived, if one looks upon $\ln 1/x$ as an additional enhancement, and selects specific NNLL contributions $\cO{\as\ln 1/x}$ in each order of the pQCD expansion. Within this approach approximate analytic expressions for the inclusive {\em energy}\/ spectrum of final particles has been derived.
In this letter we extend the analysis of \cite{GDKT} to the case of the {\em pseudorapidity}\/ (angular) distribution of particles produced in small-$x$ DIS processes.

We will treat the process in the Breit reference frame ($\mu=0$, $-q^2=Q^2$, $x=Q/2P\ll 1$). In this frame proton fragmentation occupies positive pseudorapidities (target fragmentation region)
\[
    \eta \>=\> - \ln \frac{\vartheta}{2},
\]
with $\vartheta$ the particle production angle with respect to the proton direction.
The struck quark jet populates the region $\eta <0$ (current fragmentation region).

\begin{figure}[h]
\begin{minipage}{0.7\textwidth}
   \includegraphics[width=0.7\textwidth]{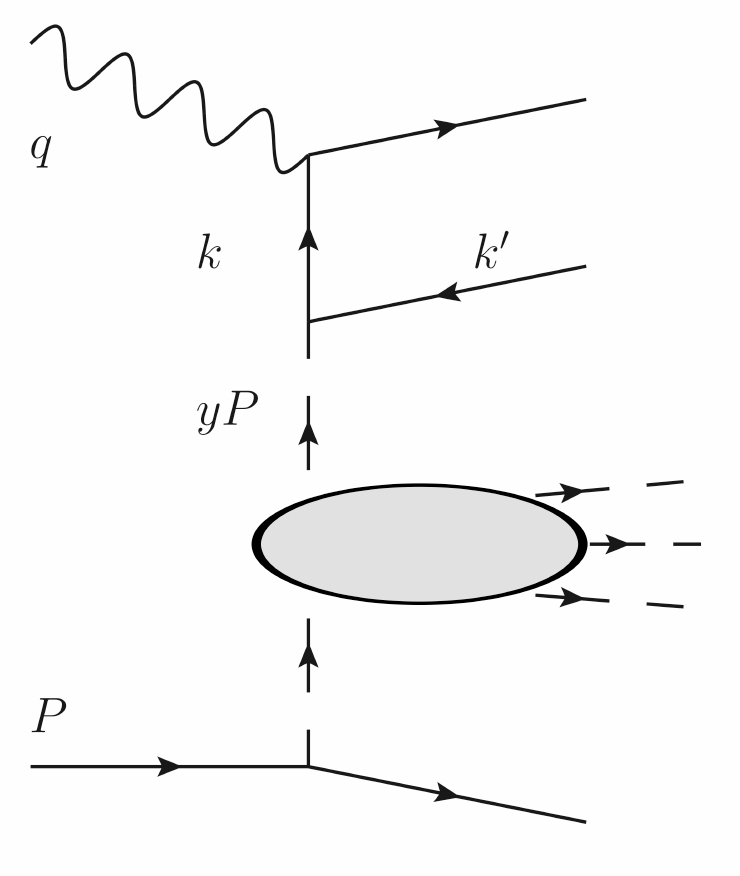}
\caption{ \label{fig:qbox} Gluon ladder attached to the quark box.}
\end{minipage}
\end{figure}

DIS cross section at small $x$ is dominated by space-like parton fluctuations that have the structure of a gluon ladder attached to the quark box as shown in Fig.~\ref{fig:qbox}.

\subsection{Quark box}

Fragmentation of the quark $k'$ in Fig.~\ref{fig:qbox} gives rise to a jet with an opening angle $\Theta' \simeq k'_t/\beta' P$.
Here $\Theta'$ is the quark production angle, and $k'_t$ and $\beta'$ --- its transverse momentum and longitudinal momentum fraction (for definition of the Sudakov decomposition of parton momenta see Appendix).
This should be taken together with radiation off the virtual quark line $k$ in the interval of Breit-frame angles
$\Theta' \le \vartheta$, thus forming a full quark jet in the target fragmentation region (proton hemisphere), similar to that in $e^+e^-\to q\bar{q}$ annihilation with invariant annihilation energy $s=Q^2$.
Similarly, the struck quark with momentum $q+k$ produces the second jet populating the current fragmentation (photon hemisphere).

Measuring preudorapidity of a final particle introduces certain competition at the level of collinear logarithmic enhancements.

Consider a gluon with positive pseudorapidity. It can belong to the jet that the quark with momentum $k'$ develops in the final state.
In this case, logarithmic integration over the relative angle $ \Theta_{{k',\,\ell}}$
between the three-momenta of the quark, $\vec{k}'$,  and that of the gluon, $\vec{\ell}$, runs in the kinematical region
\beq\label{eq:relang}
   \Theta_{{k',\,\ell}} \> \ll\> \Theta' \simeq \vartheta.
\eeq
The energy of the quark $k'$ is typically of the order of $xP=Q/2$.
Thus, logarithmic integrations over the gluon energy $\ell$ and the relative angle \eqref{eq:relang}  produce the total quark jet multiplicity factor at a hardness scale $Q\vartheta/2$,
\[
       k'_t  \>\simeq\>  k'_0\cdot \Theta' \>\sim\>  \frac{Q}{2} \, \vartheta, \quad \cN_q\left(Q\sin\frac{\vartheta}{2}\right).
\]
At the same time, the fact that the quark transverse momentum is fixed, corresponds to taking logarithmic derivative of the quark pdf $D_h^q\left(x;\mu^2 \right) $.
This gives rise to a contribution
\begin{subequations}\label{eq:qcontr12}
\beq\label{eq:qcontr1}
    \cN_q\left(Q\sin \frac{\vartheta}{2}\right) \cdot \frac{d}{d\eta}  D_h^q\left(x; Q^2\sin^2 \frac{\vartheta}{2} \right)  .
\eeq
Another logarithmic enhancement may originate from the integral over transverse momentum of the quark in the alternative region of production angles, namely
\[
     \Theta' \> \ll\> \vartheta \simeq  \Theta_{{k',\,\ell}} .
\]
This integration gives rise to the quark pdf at the same scale $Q\vartheta/2$,
while the multiplicity flow at a fixed angle $\vartheta$ is described by the derivative:
\beq\label{eq:qcontr2}
  D_h^q \left(x; Q^2\sin^2 \frac{\vartheta}2\right)    \cdot \frac{d}{d\eta}  \cN_q\left(Q\sin \frac{\vartheta}{2}\right)   .
\eeq
\end{subequations}
 The two contributions \eqref{eq:qcontr12} combine into
\beq
  \frac{d}{d\eta} \bigg[   D_h^q\left(x; Q^2\sin^2\frac\vartheta 2 \right)   \,  \cN_q\left(Q\sin \frac{\vartheta}{2}\right)   \bigg]  \qquad \eta>0.
\eeq
Analogous consideration applies to the radiation in the current fragmentation region, $\eta<0$ (with replacement of $\vartheta$ by $\pi - \vartheta$).

Finally, for entire pseudorapidity region we get an elegant expression
\beq\label{eq:qbox}
   D_h^q(x; Q^2) \,\frac{dn^{(1)}}{d\eta} =   \frac{d}{d\eta} \bigg[
    D_h^q\left(x; Q^2\sin^2\frac\vartheta 2 \right)   \cN_q\left(Q\sin \frac{\vartheta}{2}\right)
+  D_h^q\left(x; Q^2\cos^2\frac\vartheta 2 \right)  \cN_q\left(Q\cos \frac{\vartheta}{2}\right)
   \bigg] .
\eeq
Integrating over pseudorapidity,  for accompanying particle multiplicity due to radiation off the quark box we obtain
\[
    n^{(1)} \>=\>  2\, \cN_q(Q) ,
\]
which expression coincides with the mean multiplicity in $e^+e^-\to q\bar{q}$  with invariant annihilation energy $s=Q^2$.

\subsection{Soft $t$-channel radiation}

If the quark box particle production is similar to that in $e^+e^-\to q\bar{q}$,
in small-$x$ DIS there is an additional essential source of final particles that mimics a {\em gluon}\/ jet.
It is due to coherent radiation of soft gluons $\ell$ with longitudinal momenta $\beta_\ell < x$,
and emission angles {\em smaller}\/ that the production angle of the quark $k'$:
\beq\label{eq:tchanang}
    \vartheta \> <  \Theta'.
\eeq
Such soft gluons originate from coherent radiation off the $s$-channel partons at an angle {\em larger}\/ than their production angles.
These are the partons (predominantly gluons) that are produced at an early stage of the parton system evolution, ``below'' the quark box in Fig.~\ref{fig:qbox}.
Intensity of this coherent radiation is proportional to the color charge of the $t$-channel exchange, $N_c$.

Substituting, as before,  $xP=Q/2$ for the quark energy, the inequality \eqref{eq:tchanang} translates into an upper limit
for the quark transverse momentum in the box:
\beq\label{eq:tchankt}
   k_t^2 \approx k_t'^2 \approx (\beta' P \cdot \Theta')^2 \>\simeq\> (Q\Theta'/2)^2 \> > \> (Q\vartheta/2)^2 .
\eeq
Integration over $k_t^2$ under this condition yields
\beq
   D_h^q \bigg(x; Q^2 \bigg) - D_h^q \left(x; Q^2 \sin^2 \frac\vartheta 2 \right), \quad 0< \vartheta \le {\pi},
\eeq
where we have replaced $\vartheta/2$ by $\sin\vartheta/2$ to assure a smooth transition with the region of finite, $\eta=\cO{1}$, and negative rapidities (large angles $\vartheta>1$ where the collinear approximation is not applicable).
As a result, the second contribution due to soft $t$-channel radiation takes the form
\beq\label{eq:tchann}
   D_h^q(x; Q^2) \,\frac{dn^{(2)}}{d\eta} \>=\>
 \left[ D_h^q \bigg(x; Q^2 \bigg) - D_h^q \left(x; Q^2 \sin^2 \frac\vartheta 2 \right) \right]
   \cdot  \frac{d}{d\eta}  \cN_g\left(Q\sin \frac{\vartheta}{2}\right)   .
\eeq

\subsection{Fragmentation of structural gluons}

The soft $t$-channel radiation has to be combined with the fragmentation of the final state gluons that participate in formation of the quark pdf. We shall call them ``structural gluons''.

\begin{figure}[h]
\begin{minipage}{0.7\textwidth}
   \includegraphics[width=0.7\textwidth]{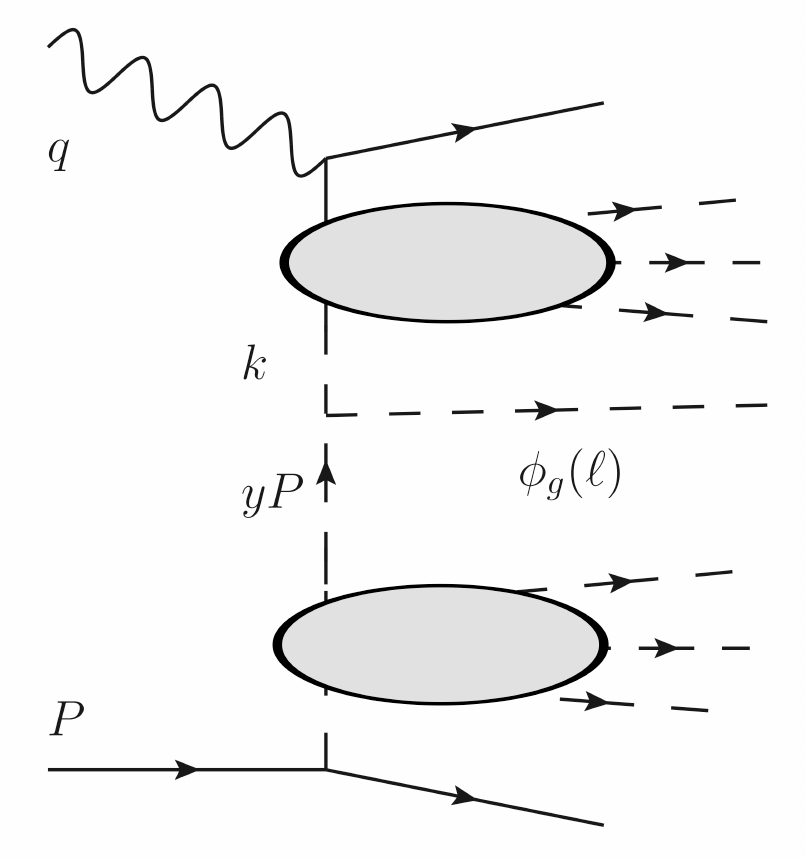}
   \caption{\label{fig:ginside} Fragmentation of a structural gluon (ladder rung).}
\end{minipage}
\end{figure}

An inclusive gluon production cross section displayed in Fig.~\ref{fig:ginside} is given by a simple convolution of two parton distributions:
\beq\label{eq:str9}
   \int_{\mu^2}^{Q^2} \frac{dk_t^2}{k_t^2}\,\bas(k_t^2)  \int_x^1 \frac{dy}{y} \,  D_h^g\bigg({y}; k_t^2\bigg)
  \bigg[  \int_{x/y}^1 \frac{dz}{z}\, \Phi_g^{g}(z) \,  D_g^q\left(\frac{x}{zy}; Q^2, k_t^2\right) \bigg]  \cdot   \phi_g(\ell).
\eeq
Here $\bas$ is conveniently normalized coupling constant (see Appendix, Eq.~\eqref{eq:basdef}), and $\Phi_g^{g}(z)$ --- the DGLAP $g\to g$ splitting function
\eqref{eq:gg}.

The first distribution $D_h^g$ in \eqref{eq:str9} is the customary gluon pdf.
It stands for the probability to find, at a certain intermediate virtuality scale $k_t^2 > \mu^2$, a gluon with the longitudinal momentum fraction $y$ inside the target hadron, with  $\mu$ the transverse momentum scale above which the pQCD approach can be applied.
The second distribution $D_g^q$ is a fundamental solution of  the system of DGLAP evolution equations.
It describes further evolution of the parton system, starting off from the gluon with the longitudinal momentum $\beta_k=z\cdot y$ (and an initial virtuality scale $k_t^2$) up to the hit quark $x$ ($Q^2$).
Finally, the factor $\phi_g(\ell)$ encodes information about fragmentation of the gluon $\ell$ and depends on the observable under consideration.

By virtue of the evolution equation for parton distributions, the $z$-convolution in the square brackets can be cast as {\em derivative}\/ over the virtuality scale:
\beq\label{eq:trickder}
   \bigg[  \int_{x/y}^1 \frac{dz}{z}\, \Phi_g^{g}(z) \, D_g^q\left(\frac{x}{zy}; Q^2, k_t^2\right) \bigg]
   \>=\> - \left[  \frac{d}{d \xi_k} D_g^q\left(\frac{x}{y}; Q^2, k_t^2\right)  \right]^{(*)} ,
\eeq
with $\xi_k$ the {\em evolution time}\/ parameter \eqref{eq:xidef}.

The quark distribution in the gluon, $D_g^q$, contains the Born contribution in which the target gluon coverts directly into a $q\bar{q}$ pair without producing any $s$-channel gluons (quark box graph). Upon differentiation over the lower scale $\xi_k$, this contribution would have produced  the $\Phi_g^q$  kernel instead of the desired gluon--gluon splitting $\Phi_g^g$. The superscript $(*)$ stands as a reminder that the quark box term is subtracted from the derivative of $D_g^q$ in \eqref{eq:trickder}.

\subsubsection{Anomalous contribution}

\begin{figure}[h]
   \includegraphics[width=0.5\textwidth]{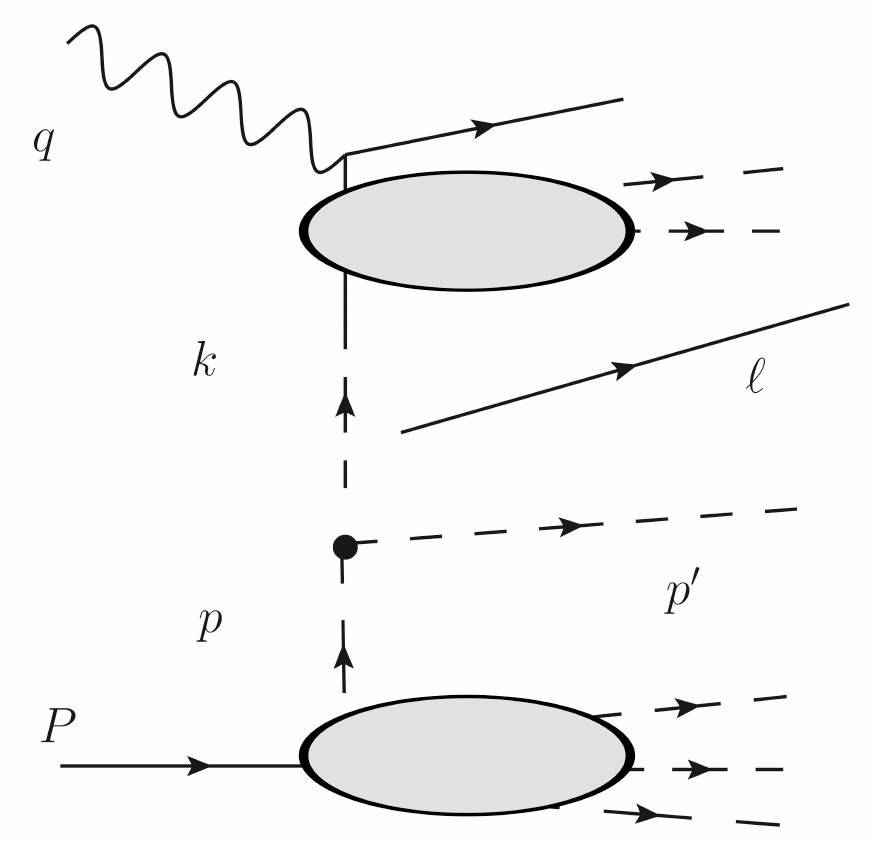}
   \caption{\label{fig:anom} ``Anomalous'' contribution $\beta_k < \beta_\ell < \beta_p$.}
\end{figure}

In the kinematical region $x\ll 1$ structural gluons are, so to say, over-ordered.
Indeed, climbing up the ladder, the longitudinal momenta of produced partons are decreasing, while their transverse momenta are strongly increasing.
As a result, the {\em emission angles}\/ are ``double ordered'', and the gluons get separated by large rapidity intervals.
When invariant pair energy $\hat{s}$ between neighboring structural gluons becomes large, the $t$-channel gluon exchange {\em reggeizes}.  This means that the {\em elastic}\/ gluon exchange amplitude acquires a suppression factor
\beq
   \left( \hat{s} \right)^{\alpha_g(t)-1}, \quad t = -k_t^2; \qquad
   \alpha_g(-\kappa^2) -1 \>\simeq\> - \int_{\mu^2}^{\kappa^2} \frac{dq^2}{q^2} \, \bas(q^2) ,
\eeq
with $\alpha_g(t)$ denoting the gluon Regge trajectory.
Physical meaning of the reggeization --- suppression of the elastic amplitude due to vetoing particle production inside a large rapidity gap (the ``fifth form factor'', \cite{FifthFF}).
In the inclusive cross section (pdf) this suppression is compensated, once again, by radiation of real gluons.
In this case the gluons $\ell$ in Fig.~\ref{fig:anom} are ``soft'' and ``hard'' at the same time.
Namely, soft with respect to the structural gluons of preceding generations, but more energetic than the exchange line:
\beq\label{eq:betarestr}
  \beta_k \>\ll\> \beta_\ell \>\ll\> \beta_p .
\eeq
In \cite{GDKT1} such gluons were referred to as ``anomalous''.
In fact, this is nothing but radiation that compensates the so-called non-Sudakov form factor suppression in the language of the CCFM scheme of generating DIS events \cite{CCFM}.
The origin of these gluons --- coherent large-angle radiation off the external lines {\em preceding}\/ the given cell.
Their transverse momenta are {\em smaller}\/ than those flowing through the cell:
\beq\label{eq:ktrestr}
    \ell_t^2 \> \ll \> k_t^2 \,\simeq p_t'^2.
\eeq
The anomalous contribution reads
\beq\label{eq:anom1}
 \int_{\ell_t^2}^{Q^2} \frac{dk_t^2}{k_t^2} \bas(k_t^2)\,
 \int^{1}_{\beta_\ell}  \frac{d\beta_p}{\beta_p}\,   D_h^g\bigg( \beta_p; k_t^2 \bigg)
 \int^{\beta_\ell}_{x}  \frac{d\beta_k}{\beta_k}\,
  \Phi_g^{g}\left(\frac{\beta_k}{\beta_p}\right)\,
 D_g^q\bigg( \frac{x}{\beta_k}; Q^2, k_t^2 \bigg).
\eeq
It resembles the standard gluon ladder but with additional kinematical restrictions imposed on the parton energies \eqref{eq:betarestr} and transverse momenta \eqref{eq:ktrestr} due to the presence of the anomalous gluon~$\ell$.

In the small-$x$ kinematics, the gluon splitting function can be approximated essentially as  $\Phi_g^g(z)\simeq 1/z$. This allows us to substitute
\beq\label{eq:PPtrick}
      \Phi_g^{g}\left(\frac{\beta_k}{\beta_p}\right) \>\Longrightarrow\>  \Phi_g^{g}\left(\frac{\beta_k}{\beta_\ell}\right)\cdot  \Phi_g^{g}\left(\frac{\beta_\ell}{\beta_p}\right)
\eeq
and to use twice the evolution equation for parton distributions, to rewrite \eqref{eq:anom1} in a compact form
\beq\label{eq:anom3}
	\int_{\xi_\ell}^{\xi_Q} d\xi_k\,
	\left[ \frac{d}{d\xi_k} \>  D_h^g\bigg( \beta_\ell; k_t^2 \bigg) \right] \cdot
	\left[  - \frac{d}{d\xi_k} \>   D_g^q\bigg( \frac{x}{\beta_\ell}; Q^2, k_t^2 \bigg) \right]^{(*)} .
\eeq
Now we can simplify the $k_t^2$ integral in \eqref{eq:anom3}. Integration by parts yields
\beq\label{eq:anom_bp}
\begin{split}
 &   \int_{\xi_\ell}^{\xi_Q} d\xi_k\,  D_h^g\bigg( \beta_\ell; k_t^2  \bigg)  \>
 \frac{d^2}{d\xi_k^2} \> D_g^q\bigg( \frac{x}{\beta_\ell}; Q^2, k_t^2 \bigg) \cr
 & +  D_h^g\bigg( \beta_\ell; \ell_t^2  \bigg)  \>
\left[  \frac{d}{d\xi_\ell} \> D_g^q\bigg( \frac{x}{\beta_\ell}; Q^2, \ell_t^2 \bigg)  \right]^{(*)}.
  \end{split}
\eeq
The second line represents the surface term ($k_t^2=\ell_t^2$) which cancels with the structural contribution~\eqref{eq:trickder}.
(The second surface term, $k_t^2=Q^2$, vanishes.)

Adding together \eqref{eq:str9}, \eqref{eq:trickder} and \eqref{eq:anom_bp} gives
\beq\label{eq:structplusanom2}
  \int_{\ell_t^2}^{Q^2} \frac{dk_t^2}{k_t^2} \, \bas(k_t^2) \,  D_h^g\bigg( \beta_\ell; k_t^2 \bigg)  \>
 \frac{d^2}{d\xi_Q^2} \> D_g^q\bigg( \frac{x}{\beta_\ell}; Q^2, k_t^2 \bigg) .
\eeq
Here we have replaced the second derivative over $\xi_\ell$ by that over $\xi_Q$ since the parton distribution depends on the difference of evolution times, $\xi_Q-\xi_\ell$.

The expression \eqref{eq:structplusanom2} has to be supplied with the final state factor $\phi_g(\ell)$ that describes fragmentation of the gluon $\ell$.
This factor depends on the observable under consideration. In our case of multiplicity flow at a given pseudorapidity it reduces to the mean parton multiplicity in the gluon jet $\cN_g$ with the hardness parameter $\ell_t^2$.
The ratio of secondary parton multiplicities in gluon and quark jets is known to next-to-next-to leading order \cite{GaffMuel}, and numerically is close to the ratio
of color factors $N_c/C_F=9/4$.

For the sum of the structural and anomalous contributions we finally obtain
\beq\label{eq:anomfin}
  D_h^q(x;Q^2) \cdot \frac{dn^{(3)}}{d\eta} \>=\>  2 \int_x^1 \frac{d\beta_\ell}{\beta_\ell} \bas(\ell_t^2)
   \, \cN_g(\ell_t^2) \cdot
     \int_{\ell_t^2}^{Q^2} \frac{dk_t^2}{k_t^2} \, \bas(k_t^2) \,  D_h^g\bigg( \beta_\ell; k_t^2 \bigg)  \>
 \frac{d^2}{d\xi_Q^2} \> D_g^q\bigg( \frac{x}{\beta_\ell}; Q^2, k_t^2 \bigg) .
\eeq
Here the gluon radiation angle $\vartheta$ stays fixed, while its energy $\beta_\ell$ is integrated over.
The virtuality scale $\ell_t^2$ changes together with $\beta_\ell$:
\[
    \ell_t = \beta_\ell P \cdot  \vartheta = \frac{\beta_\ell}{x} \cdot xP \vartheta
    \>\simeq\>  \frac{\beta_\ell}{x} \cdot Q \sin \frac{\vartheta}{2}.
\]
Note that for sufficiently large positive pseudorapidities  $\eta$ (small emission angles $\vartheta$),
the transverse momentum $\ell_t$ of the radiated gluon inside the integration region may become smaller than a critical value $\mu$ below which the pQCD approach is no longer applicable (the coupling may hit the ``Landau pole'').
A transverse momentum cutoff  $\ell_t>\mu$ has to be introduced, and the third component \eqref{eq:anomfin}
becomes collinear sensitive.
This happens for $\eta \ge \log(Q/2\mu)$.

We chose a small value of $\mu$ in order to put maximal responsibility for particle production on the pQCD dynamics.
We set $\mu^2 = 0.2 \GeV^2$ which value corresponds to
the initial scale of GRV parton distributions~\cite{GRV}.

Since gluons with small transverse momentum $\ell_t\sim \mu$ do not cascade, collinear sensitivity of the answer turns out to be moderate.
An uncertainty due to variation of  the collinear cutoff is restricted to a narrow interval
of 0.5$\div$1 units in rapidity around $\eta= \log(Q/2\mu)$, amounts to several percent and decreases with increase of the hardness of the process $Q^2$.

\subsubsection{Analytic estimate of the magnitude of the third component}

In oder to estimate relative weight of  the third contribution, let us consider its share in the total particle yield.
Introducing an integral over the full rapidity range ($\eta>0$) unties the $\beta_\ell$ and $\ell_t$ integrations by making them independent.
Then, the convolution of the two successive parton distributions over the longitudinal momentum fraction $\beta_\ell$, by virtue of the completeness relation, yields
\beq
  \int_x^1 \frac{d\beta_\ell}{\beta_\ell}  \>  D_h^g\bigg( \beta_\ell; k_t^2 \bigg)  \cdot  D_g^q\bigg( \frac{x}{\beta_\ell}; Q^2, k_t^2 \bigg)
  \>\simeq \>     D_h^q\big( x; Q^2) .
\eeq
Since the answer does not depend on an intermediate scale $k_t^2$, one immediately arrives at
\beq
    n^{(3)} \>=\>  \int_{\mu^2}^{Q^2} \frac{d\ell_t^2}{\ell_t^2}\, \bas(\ell_t^2)   \, \cN_g(\ell_t^2) \cdot (\xi_Q-\xi_\ell)
\times \left[  \left. \frac{d^2 D_h^q\big({x}; Q^2 \big)}{d\xi_Q^2} \right/ D_h^q(x;Q^2)\right] .
\eeq
Given a sharp increase of the mean multiplicity factor $\cN_g$ with $\ell_t^2$,
\[
     \cN'/\cN \>\propto \sqrt{\bas},
\]
an estimate follows:
\beq
  \lrang{  \frac{d\ell_t^2}{\ell_t^2}\, \bas(\ell_t^2)   } \> \sim \> \lrang {\xi_Q-\xi_\ell } \>=\>\cO{\sqrt{\bas}}.
\eeq
Invoking the known enhancement of the pdf scaling violation rate at small $x$, $d/d\xi \ln D
\propto \sqrt{\ln x^{-1}}$, one finally obtains
\beq
 \frac{n^{(3)}}{\cN_g} \>=\> \cO{\bas \ln x^{-1}}.
\eeq
As envisaged in the introduction to this Section, this contribution is formally of NNLL nature, as it is proportional to the second power of $\sqrt{\bas}$.
However, the fact that it is enhanced by the $\ln x$ factor, makes it legitimate to keep this term while neglecting other (non-enhanced) NNLL corrections.

\section{Numerical results  \label{sec:2}}

\subsection{Ingredients}

The final result is given by the sum of three contributions:  \eqref{eq:qbox}, \eqref{eq:tchann} and \eqref{eq:anomfin}.
These formulae contain three main ingredients.
\begin{enumerate}
\item
	$\cN_g(k^2)$ --- multiplicity of partons in a gluon jet at hardness scale $k^2$.
For this function we employed the analytic expression derived in the so called Modified Leading Logarithmic Approximation (MLLA).
In our numerical calculations we used the parameter-free ``limiting spectrum'' approximation, that one obtains by setting the collinear cutoff $Q_0=\Lambda_{QCD}$.
We chose $\Lambda_{QCD}$=320 \MeV, which value provided the best fit to LEP data~\cite{Delphi}. In addition, for numerical analysis we use the leading order relation between parton multiplicities in quark and gluon jets
$\cN_q=(C_F/C_A)\cN_g$.

\item
	$D_g^q(x; Q^2,k^2)$ --- the two-scale fundamental solution of the DGLAP evolution equations that describes distribution of sea quarks in the target gluon. This distribution is calculated numerically by inverting the corresponding LLA expression in the Mellin moment space $N$, known analytically, to  the $x$-space.
\item  $D_h^q(x; k^2)$ and $D_h^g(x; k^2)$  --- quark and gluon distributions in the target hadron.
We employed the Gluck--Reya--Vogt (GRV) pdfs~\cite{GRV} at a low virtuality scale $\mu^2=0.2\,\GeV^2$, and used
 $\Lambda_{\rm \overline{MS}}=230$ \MeV\ to evolve them to arbitrary hardness scales $k^2$. This value of $\Lambda_{\rm \overline{MS}}$
matches the above value $\Lambda_{\rm QCD}$=320 \MeV\ that corresponds to the physical (``bremsstrahlung'', ``MC'') scheme for the QCD coupling~\cite{MCW}.
\end{enumerate}

%
%
%
%

\subsection{Figures}

We illustrate our results with several numerical examples.

In order to demonstrate how does the pseudorapidity distribution evolves with $Q^2$ and $x$,
we present here the curves for three momentum transfers, $Q=$10, 100 and 1000 \GeV\, for two values of the Bjorken variable $x=10^{-2}$ and $x=10^{-4}$.

\begin{figure}[h]
\begin{minipage}{0.45\textwidth}
    \includegraphics[width=0.95\textwidth]{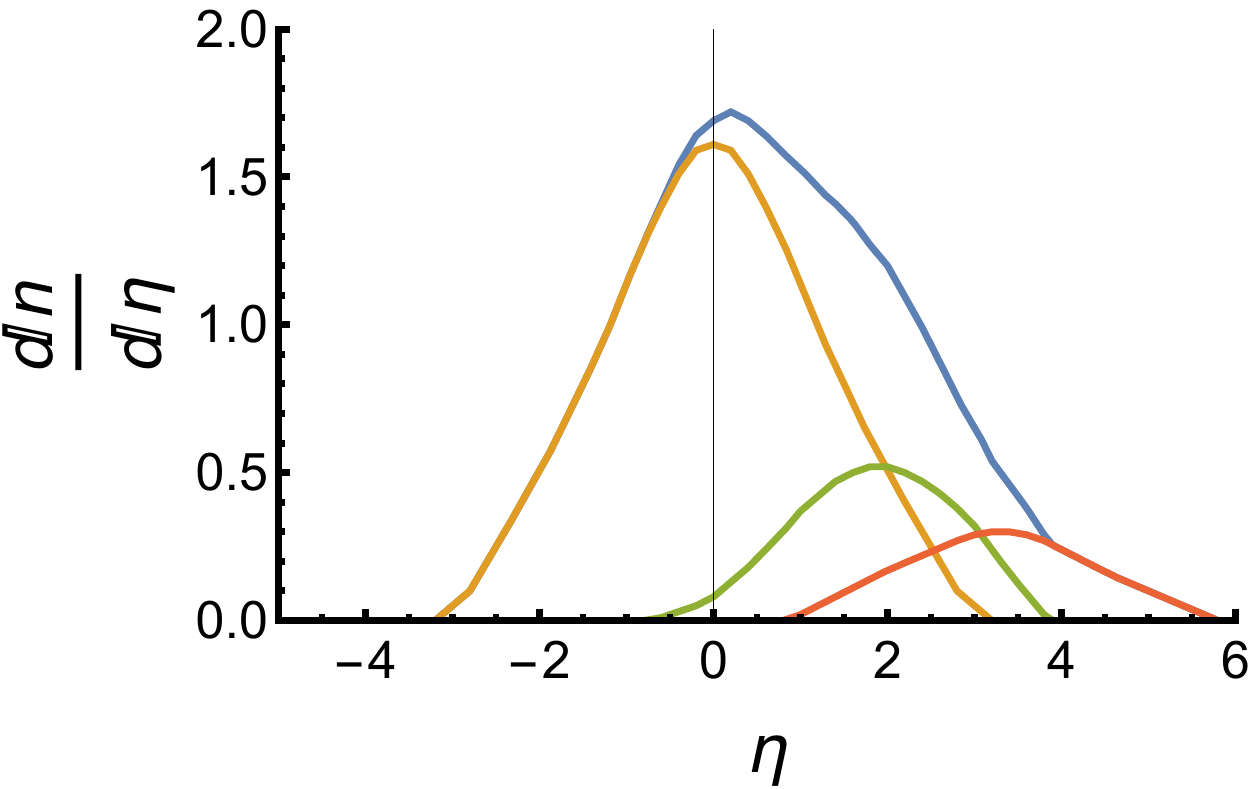}
\end{minipage}
\begin{minipage}{0.45\textwidth}
    \includegraphics[width=0.95\textwidth]{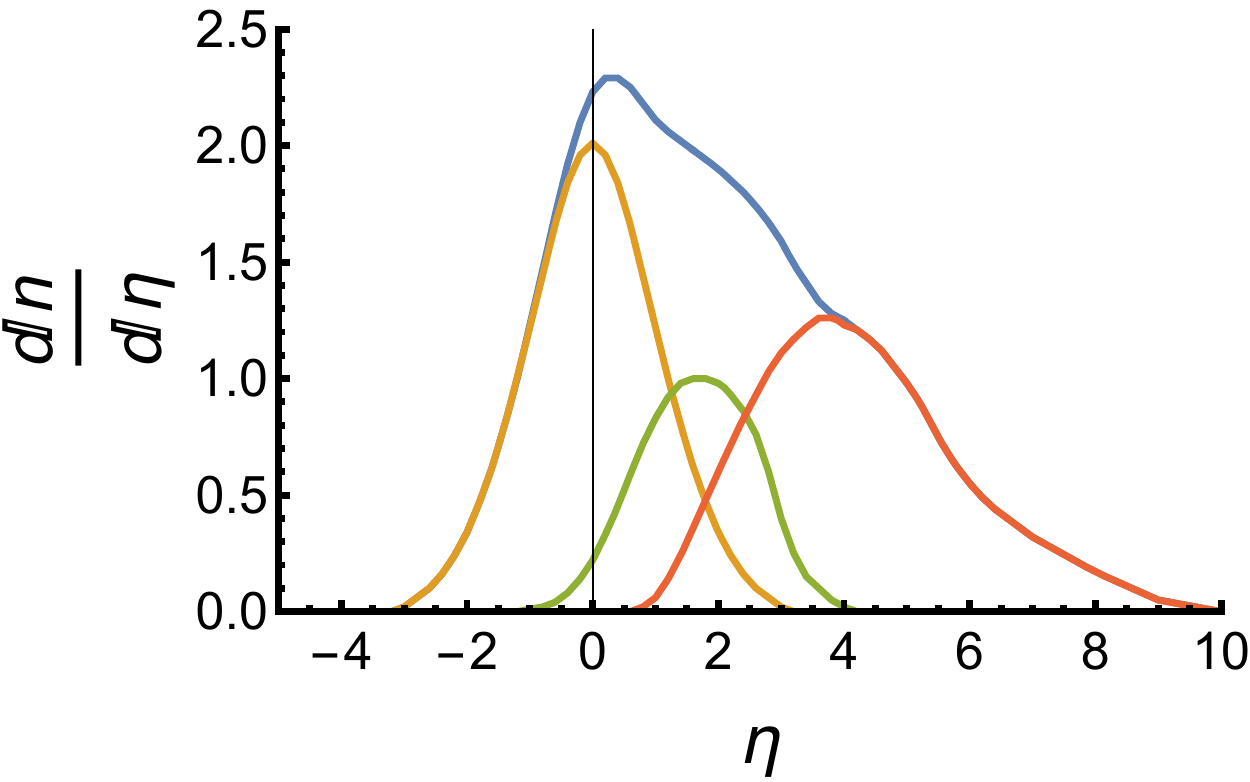}
\end{minipage}
\caption{\label{fig:1}  $Q=10\, \GeV$. Left panel $x=10^{-2}$, right panel $x=10^{-4}$.  }
\end{figure}

\begin{figure}[h]
\begin{minipage}{0.45\textwidth}
    \includegraphics[width=0.95\textwidth]{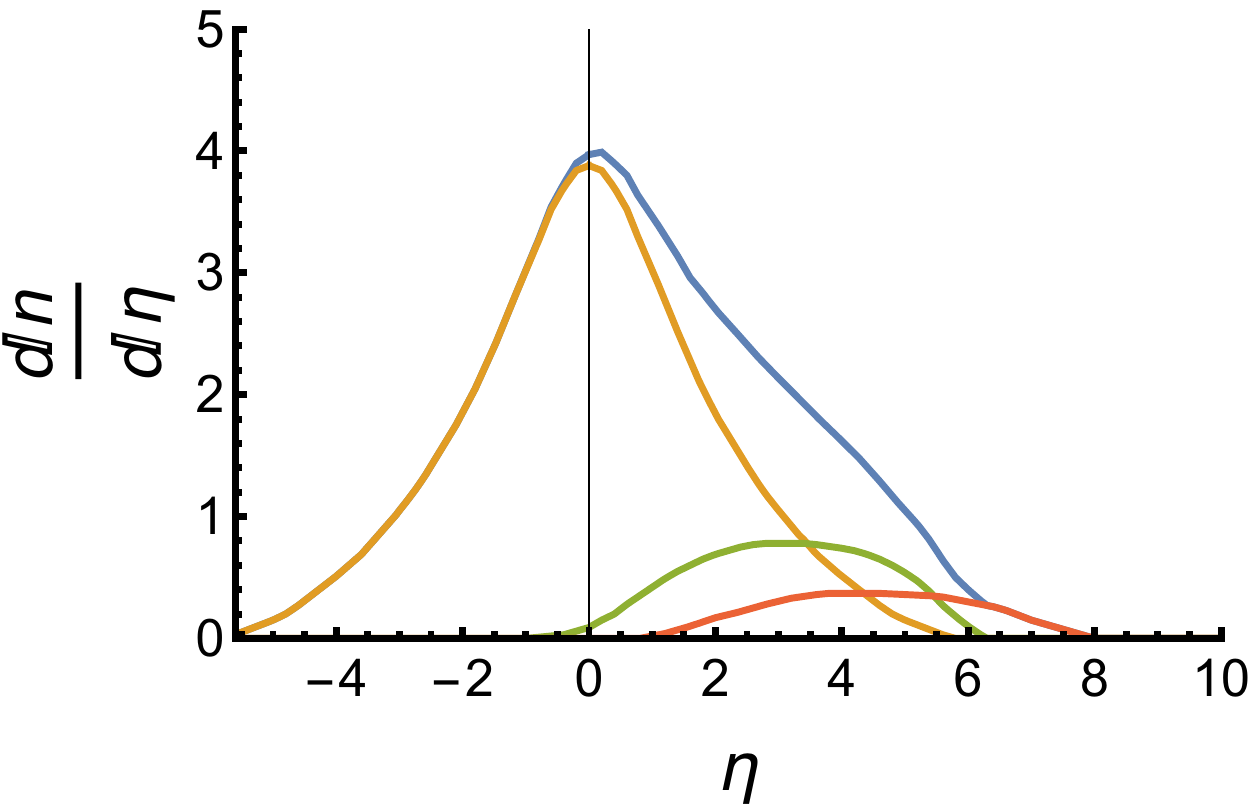}
\end{minipage}
\begin{minipage}{0.45\textwidth}
    \includegraphics[width=0.95\textwidth]{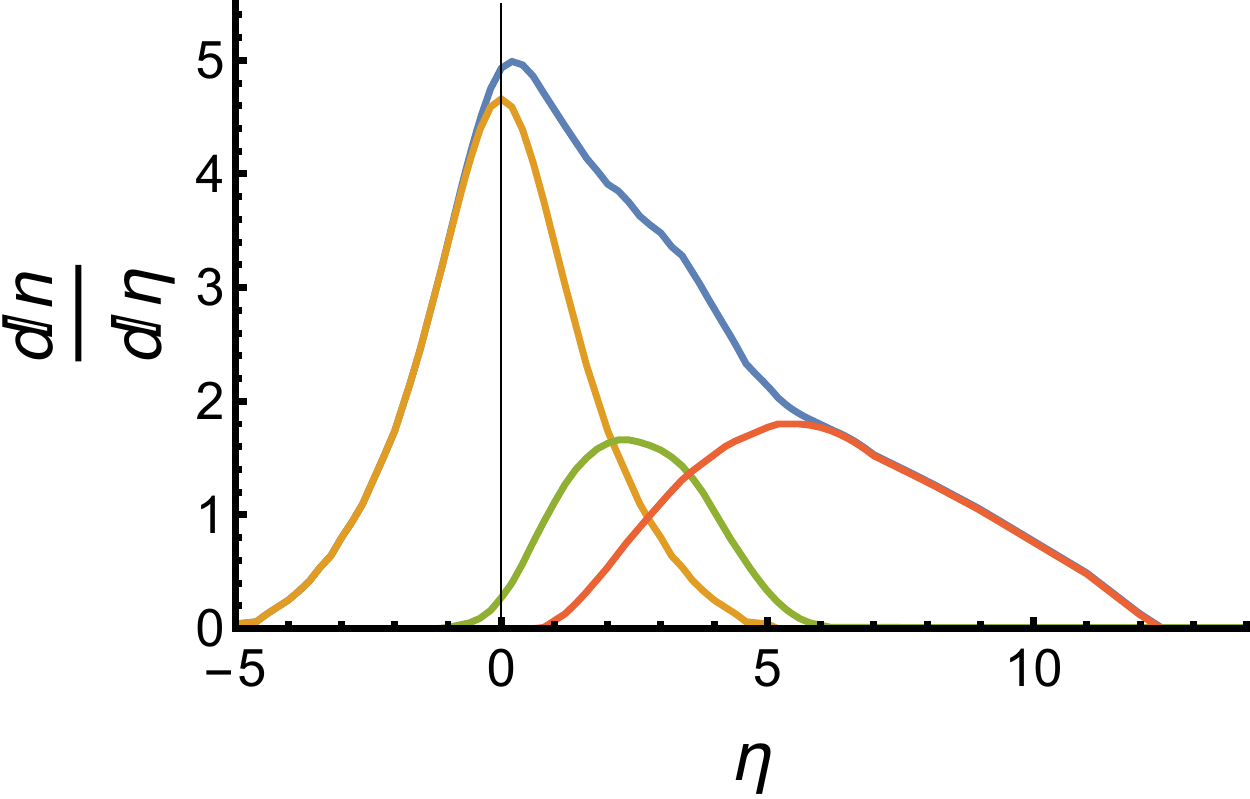}
\end{minipage}
\caption{\label{fig:2}  $Q=100\, \GeV$. Left panel $x=10^{-2}$, right panel $x=10^{-4}$.  }
\end{figure}

\begin{figure}[h]
\begin{minipage}{0.45\textwidth}
    \includegraphics[width=0.95\textwidth]{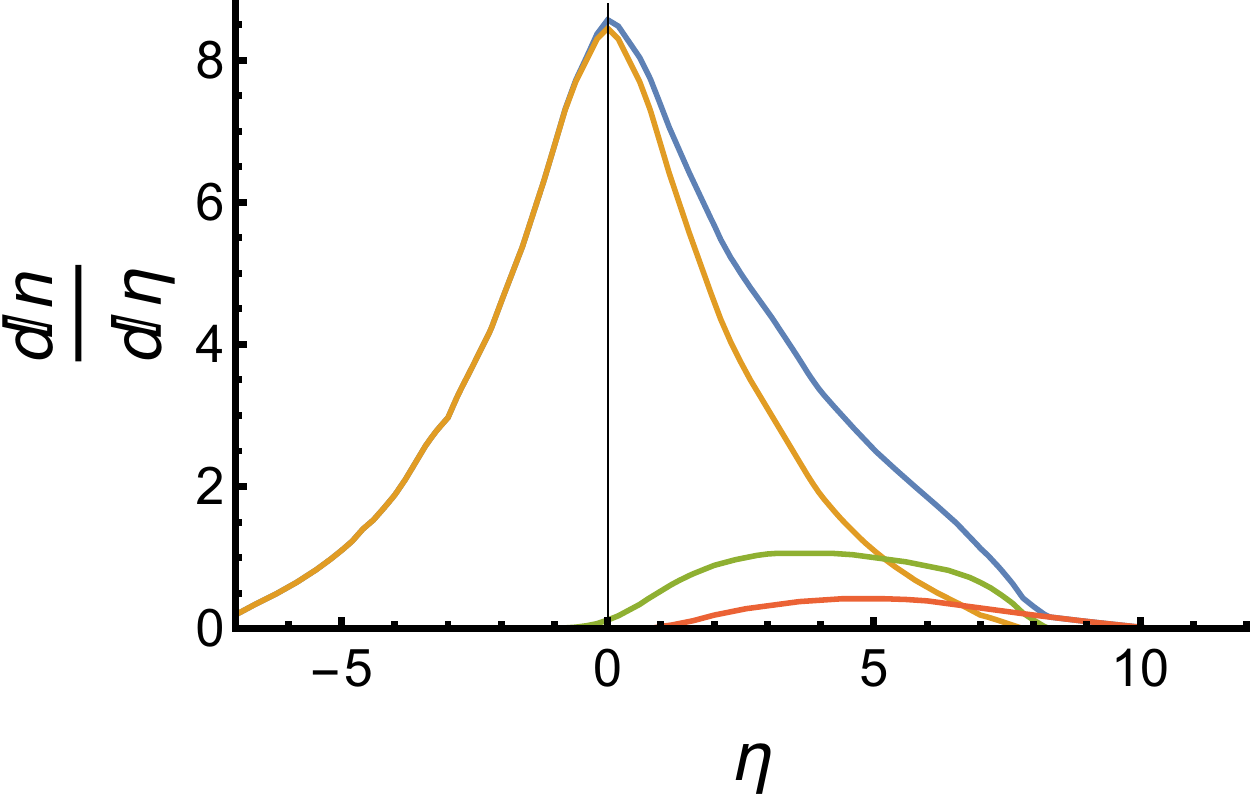}
\end{minipage}
\begin{minipage}{0.45\textwidth}
    \includegraphics[width=0.95\textwidth]{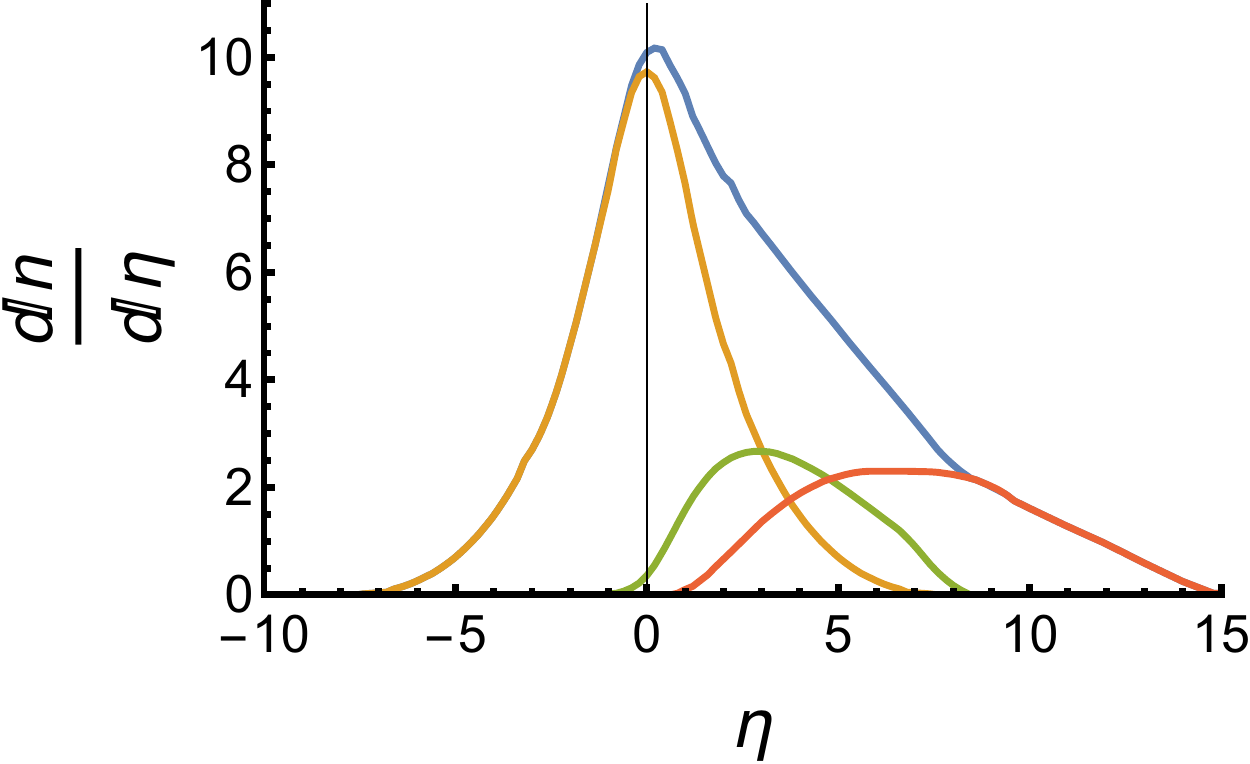}
\end{minipage}
\caption{\label{fig:3}  $Q=1000\, \GeV$. Left panel $x=10^{-2}$, right panel $x=10^{-4}$.  }
\end{figure}

The upper curve In Figs.~\ref{fig:1}--\ref{fig:3} shows the density of final state particles (charged hadrons)
as a function of  the Breit frame pseudorapidity.
It is a sum of three contributions described above.

\begin{itemize}
\item
The leftmost curve describes fragmentation of the ``quark box'' given by \eqref{eq:qbox} (contribution 1).
It is symmetric around $\eta=0$ and concentrated in the pseudorapidity interval $|{\eta}|\le  \ln(Q/\Lambda_{QCD})$.
\item
The middle curve is due to accompanying soft gluons radiation in the target fragmentation region, \eqref{eq:tchann} (contribution 2).
\item
The rightmost curve combines fragmentation of structural gluons and the anomalous contribution due to gluon reggeization, \eqref{eq:anomfin}
(contribution 3).
It extends to large positive pseudorapidities and strongly increases with $x$ decreasing.
\end{itemize}


We also calculated the pseudorapidity distribution of the final particle multiplicity flow in DIS
for  $Q^2=20\> \GeV^2$ and $x=10^{-3},10^{-4}$  typical for HERA kinematics.
Comparison of QCD expectations with the new HERA data~\cite{H1new} will be reported elsewhere.
	
	\clearpage

\section{Conclusions}

There are two sources of final state particles in DIS processes: fragmentation of partons forming pdfs, and accompanying coherent soft gluon radiation caused by $t$-channel color exchange. Both these sources are adequately embedded, in particular, in the CASCADE MC generator developed by G.~Salam and J.~Jung, which event generator incorporates both CCFM and BFKL physics \cite{Salam1999,CASCADE2001}.

Analytic pQCD analysis is an alternative to MC generation of events. In this letter we derived, in a compact analytic form, pQCD predictions for pseudorapidity distributions of final state particles produced in DIS processes at small Bjorken $x$.

 It would be highly desirable to reanalyzed the HERA data on hadron production
 \cite{H1new} as a function $dn/dy $ and compare them with QCD predictions.


It  is straightforward to generalize and apply the above analysis to the double-differential distribution of final particles in pseudorapidity and  energy.

A similar approach can be used to study the initial radiation in $pp$ collisions at LHC.

\appendix

\section{Notation}

\noindent
Sudakov decomposition:
\beq
        k \>=\> \beta_k P \>+\> k_t \>+\> \alpha_k q', \qquad q'=q+xP, \quad q'^2 =P^2 = 0.
\eeq
Coupling constant
\beq\label{eq:basdef}
  \bas \>=\> \frac{N_c\,\as}{\pi}.
\eeq
In this normalization, the parton splitting functions that appear in the text are
\begin{subequations}\label{eq:splitfunctions}
\beeq
\label{eq:gg}
 \Phi_g^g(z) \>&=&\> \frac{1-z}{z} + \frac{z}{1-z} + z(1-z) , \\
\label{eq:gq}
 \Phi_g^q(z) \>&=&\> \frac{1}{2N_c} \left[ z^2 + (1-z)^2 \right]  .
\eeeq
\end{subequations}
Evolution time parameter
\begin{subequations} \label{eq:xidef}
\beq
      d\xi_k \>\equiv\> \bas(k_t^2)\, \frac{dk_t^2}{k_t^2} .
\eeq
For one-loop coupling constant,
\beq\label{eq:xi1}
  \xi_k \equiv \xi(k^2)  \>=\>   \frac{12N_c}{11N_c -2n_f} \ln \frac1{\as(k^2)} \>+\>\mbox{const.}
\eeq
\end{subequations}


\begin{thebibliography}{99}

\bibitem{AOjet}
A.H.\ Mueller,
Phys.\ Lett.\ {104 B} (1981) {161}; \\
V.S.\ Fadin,
Yad.\ Fiz.\ 37 (1983) 408.

\bibitem{GDKT}
 L.V.\ Gribov, Yu.L.\ Dokshitzer, S.I.\ Troian and V.A.\ Khoze, \\
JETP Lett.\ {\bf 45} (1987) 515 
[Pisma Zh.\ Eksp.\ Teor.\ Fiz.\  45 (1987) 405; \\ 
  Sov.\ Phys.\ JETP 67 (1988) 1303   [Zh.\ Eksp.\ Teor.\ Fiz.\   94 (1988) 12].

\bibitem{GDKT1}
L.V.\ Gribov, Yu.L.\ Dokshitzer, V.A.\ Khoze and S.I.\ Troian,
Phys.\ Lett.\ B 202 (1988) 276.  

\bibitem{Ciafaloni}
M.\ Ciafaloni,
Nucl.\ Phys.\ B 296 (1988) 49. 


\bibitem{CCFM}
S.\ Catani, F.\ Fiorani and G.\ Marchesini,
Phys.\ Lett.\ B 234 (1990) 339;   
Nucl.\  Phys.\  B 336 (1990) 18; \\
G.\ Marchesini, Nucl. Phys. B 445 (1995) 49.


\bibitem{CDW92}
S.\ Catani, Yu.L.\ Dokshitzer and B.R.\ Webber,
Phys.\ Lett.\  B 285 (1992) 291. 

\bibitem{FifthFF}
  Yu.L.\ Dokshitzer and G.\ Marchesini,
  Phys.\ Lett.\ B  631 (2005) 118. 

\bibitem{Book}Yu.L.\ Dokshitzer, V.A.\. Khoze, A.H.\ Mueller and S.I.\ Troian,
  ``Basics of perturbative QCD,''
   Gif-sur-Yvette, France: Ed. Frontieres (1991) 274 p.


%

\bibitem{Delphi}
 P.\ Abreu {\it et al.} [DELPHI Collaboration],
   Phys.\ Lett.\ B {\bf 449} (1999) 383.

\bibitem{GRV}
  M.\ Gluck, E.\ Reya and A.\ Vogt,
  Z.\ Phys.\ C {\bf 53} (1992) 127.

\bibitem{MCW}
 S.\ Catani, B.R.\ Webber and G.\ Marchesini,
   Nucl.\ Phys.\ B {\bf 349} (1991) 635.

\bibitem{Salam1999}
G.P.\ Salam, JHEP03 (1999) 009.

\bibitem{CASCADE2001}
H.\ Jung and G.P.\ Salam,  Eur.\ Phys.\ J.\ {\bf C19} (2001) 351. 

\bibitem{H1new} C.~Alexa {\it et al.} [H1 Collaboration],
   Eur.\ Phys.\ J.\ C {\bf 73} (2013) 4,  2406.   


\bibitem{GaffMuel}
J.B.\ Gaffney and A.H.\ Mueller,
Nucl.\ Phys.\ B {\bf 250} (1985) 109.

\end{thebibliography}
\end{document}